\documentclass[english,showpacs,twocolumn,aps,reprint]{revtex4-1}

\usepackage{mathptmx}
\usepackage[T1]{fontenc}
\usepackage[latin9]{inputenc}
\usepackage{color}
\usepackage{amsmath}
\usepackage{amssymb}
\usepackage{hyperref}
\usepackage{graphics}
\usepackage{graphicx}
\usepackage{epsfig}
\usepackage{bm}
\usepackage{epstopdf}
\usepackage{mathtools}
\usepackage{soul}
\usepackage{booktabs}  
\usepackage{amsmath,amssymb}
\usepackage{geometry}
\usepackage{graphicx}
\usepackage{setspace}
\usepackage{siunitx} 
\usepackage{subfigure}  
\usepackage{xmpmulti}
\usepackage{titlesec}
\usepackage{times}
\usepackage{xcolor}
\usepackage[normalem]{ulem}

\usepackage{babel}

\begin{document}

\title{{\color{blue}Role of Magnetic Field in the Redistribution of Turbulence from Large-Scale Structures to Small-Scale Fluctuations
}}

\author{Tanmay Karmakar\textsuperscript{1,2}}
\email{tanmay.karmakar@ipr.res.in}

\author{Rosh Roy\textsuperscript{1,2}}

\author{Lavkesh Lachhvani\textsuperscript{1,2}}
\author{Raju Daniel\textsuperscript{1,2}}
\author{Bhoomi Khodiyar\textsuperscript{1,2}} 
\author{Prabal K. Chattopadhyay\textsuperscript{1,2}}
\author{Abhijit Sen\textsuperscript{1,2}}
\author{Sayak Bose\textsuperscript{3,a}}

\affiliation{\textsuperscript{1}Institute for Plasma Research, HBNI, Bhat, Gandhinagar, 382428, India} 

\affiliation{\textsuperscript{2}Homi Bhabha National Institute, Anushaktinagar, Mumbai, Maharastra 400094, India}

\affiliation{\textsuperscript{3}Columbia Astrophysics Laboratory, Columbia University, 550 West 120th Street, New York, New York 10027, USA}

\affiliation{\textsuperscript{a}Present address: Princeton Plasma Physics Laboratory, Princeton, New Jersey 08540, USA.}

\begin{abstract}
\noindent

\noindent
Magnetized plasmas with equilibrium density gradients support drift-wave turbulence, which is often regulated by self-generated zonal flows. In this work, we experimentally examine the effect of increasing the magnetic field on turbulence characteristics in a linear plasma device. As the magnetic field is increased from 600 to 1000~G, zonal flow is suppressed while the mean flow increases. Spectral analysis of density and potential fluctuations shows a redistribution of power from low-frequency (0.1-1 kHz) to high-frequency (1-300 kHz) components, along with an increase in the spectral slope and the ratio $P_{\mathrm{HF}}/P_{\mathrm{LF}}$. This change is linked to a reduction in Reynolds stress due to the loss of correlation between radial and poloidal velocity fluctuations, which possibly weakens the drive for zonal flow generation. Similar behavior is observed near the peak gradient region, also indicating its global nature. The present results suggest a transition from a zonal-flow-dominated regime to a state dominated by smaller-scale fluctuations, possibly influenced by mean flow shear. These findings highlight how the magnetic field redistributes spectral energy across frequency scales in drift-wave turbulent plasmas.
\end{abstract}
\maketitle 

\section{Introduction}\label{sec1} 
\noindent
In magnetized plasmas with finite pressure gradients, drift waves arise due to electron diamagnetic drift and can become unstable in the presence of dissipation \cite{horton1999drift, schroder2005drift, marshall1986collisional}. These waves grow from small perturbations in density and potential and can reach finite amplitudes through the fluctuating $\mathbf{E}\times\mathbf{B}$ drift. As their amplitude increases, nonlinear interactions among different modes lead to energy transfer across scales through wave-wave coupling processes that satisfy frequency and wavenumber matching conditions. Under sufficiently strong drive, this process leads to a broad spectrum of fluctuations, marking the transition from coherent drift waves to fully developed drift-wave turbulence \cite{burin_2005transition, tynan_2006observation}. Such turbulence is responsible for enhanced cross-field transport of particles, momentum, and energy. This remains a central problem in plasma physics, with strong relevance to space and astrophysical plasmas \cite{howes2008kinetic, chen2016recent} as well as to magnetic confinement fusion \cite{carreras2002progress}, where it often limits confinement performance. A key feature of drift-wave turbulence is the nonlinear interaction among multiple modes, which can lead to an inverse cascade, transferring energy from small-scale fluctuations to large-scale, low-frequency coherent structures known as zonal flows \cite{diamond2005zonal, itoh2006physics}. These azimuthally symmetric shear flows are generated by the radial variation of Reynolds stress and act back on the turbulence by shearing and decorrelating turbulent eddies, thereby regulating transport        \cite{Terry_2000suppression}. This mutual interaction between turbulence and zonal flows, often described as a predator$-$prey process, has been widely studied in theoretical, numerical, and experimental works \cite{diamond2005zonal, fujisawa_2008review,fujisawa_2004identification, itoh2006physics, BDT_Theory_1990influence}. However, there are situations where this regulation mechanism weakens, leading to a reorganization of turbulence across scales, and the conditions under which this occurs are not yet fully understood. In particular, the role of the externally applied magnetic field in determining the turbulence scale, modifying the balance between turbulence drive and nonlinear interactions, and governing the re-organisation of spectral power across different turbulence scales remains an open question. An increase in magnetic field can alter turbulence characteristics by changing fluctuation scales, turbulent de-correlation rates, and nonlinear coupling among scales, which may redistribute fluctuation energy from large-scale coherent structures to smaller-scale broadband fluctuations. Such spectral changes can also reduce the efficiency of energy transfer to zonal flows, thereby modifying the turbulence regulation mechanism \cite{carter2009_modifications, floriani2013self}. While these processes are highly relevant to fusion plasmas, their direct investigation in tokamak geometry is often challenging due to the complex magnetic configuration, the simultaneous presence of multiple interacting phenomena that make it difficult to isolate the root cause. In contrast, linear magnetized plasma devices provide a simpler and more controlled steady state environment, where key parameters such as magnetic field and profile gradients can be varied independently, allowing a clearer understanding of the fundamental physics of turbulence dynamics. In this work, we experimentally investigate the evolution of drift-wave turbulence with increasing magnetic field in a linear magnetized plasma. We observe a systematic reduction in low-frequency  fluctuation power and a corresponding increase in high-frequency components, indicating a redistribution of energy toward smaller scales. At the same time, the zonal flow amplitude decreases, while there is an enhancement of the spectral power in the high-frequency band. Analysis of Reynolds stress shows a reduction in correlated velocity fluctuations, suggesting weaker momentum transfer responsible for zonal flow generation. These results indicate a transition from a zonal-flow-regulated regime to a state dominated by mean shear, where turbulence is not suppressed but reorganized across smaller scales. The measurements are performed using Langmuir probe diagnostics to obtain ion density and floating potential fluctuations, from which spectral properties, fluctuations in velocity components, and nonlinear interactions are analyzed over a range of magnetic field strengths. These findings provide new insights into how the magnetic field influences turbulence and reveals a regime where turbulence is reorganized through a redistribution of energy from low-frequency, large-scale structures to smaller-scale, high-frequency fluctuations, indicating a modification of the usual self-regulation process. \\

\noindent
The paper is organized as follows. Section~II describes the experimental apparatus, including the probe diagnostics and data analysis methods employed in this study. Section~III presents the mean radial profiles for successive values of $B_m$. In Section~IV, the evolution of turbulence spectra is examined in detail. This section is divided into subsections, discussing the emergence of low-frequency fluctuations, followed by high-frequency fluctuations, and finally the spectral reorganization of turbulence. The paper concludes with a discussion and summary of the main results in Section~V.

\section{Experimental apparatus}

\begin{figure*}[!hbt]
\centering
\includegraphics[width=\linewidth]{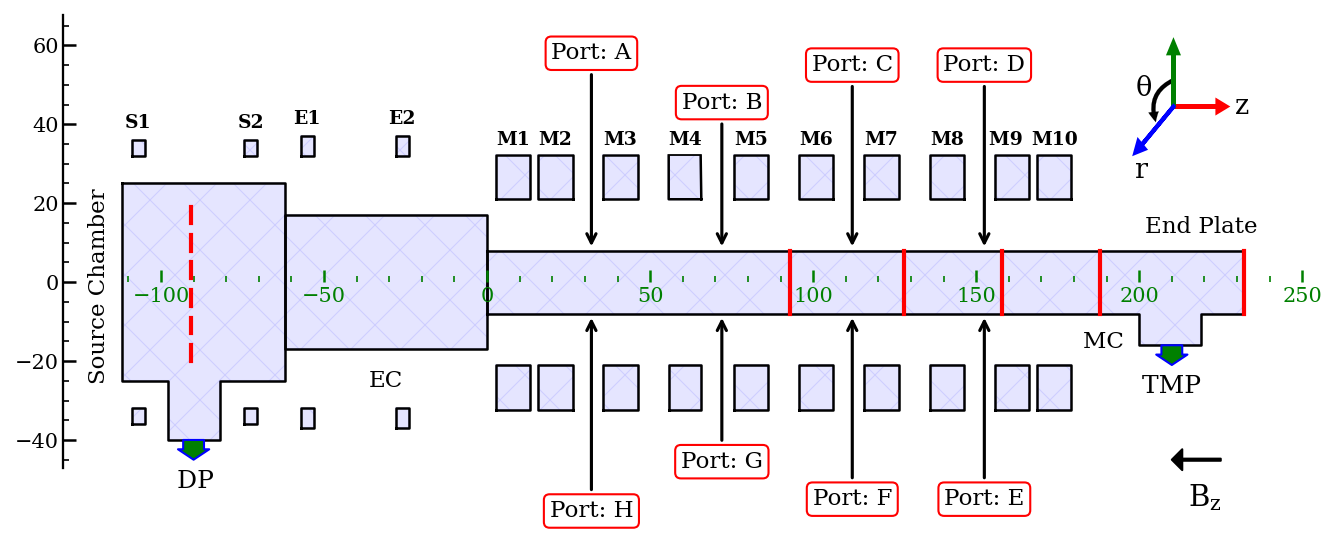} 
\caption{Schematic of the IMPED experimental setup (scale in cm). The six dashed vertical red lines at $-90$ cm indicate the filaments.}    
\label{fig:imped_device_schematic}  
\end{figure*}

\begin{figure*}[!hbt]
\centering
\includegraphics[width=0.8\linewidth]{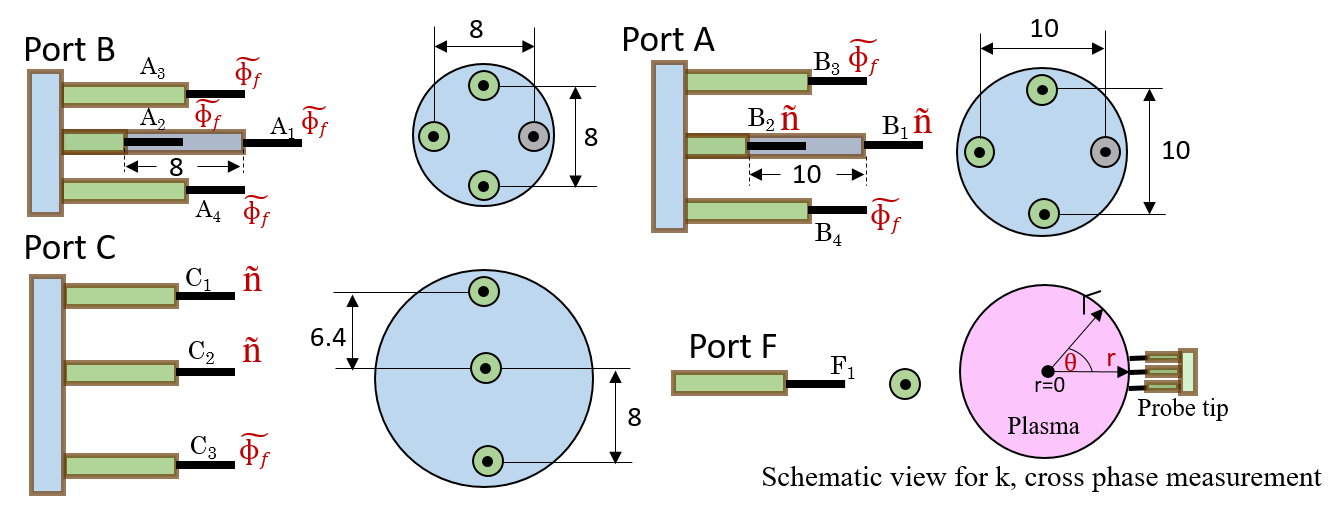} 
\caption{Schematic of the probe arrangement for experimental measurements. The probe tips are inserted through different radial ports in the $r$--$\theta$ plane and are oriented perpendicular to the axial magnetic field ($B_z$).} 
\label{fig:probe_configaration}  
\end{figure*}

\paragraph{\textbf{Device Description and Operating Conditions:}} The experiments were performed using the Inverse Mirror Plasma Experimental Device (IMPED) \cite{bose_2015inverse, roy2025_experimental, Roy_RT-KH_10.1063/5.0291547, karmakar2026zonal}, as illustrated in Fig.~\ref{fig:imped_device_schematic}. The device provides a flexible operational regime over a wide range of plasma parameters. The vacuum chamber is evacuated using a turbo-molecular pump (TMP) with a pumping speed of \SI{2200}{\liter\per\second}, allowing a base pressure of \SI{2e-6}{\milli\bar} to be achieved prior to plasma operation. Plasma is produced using an array of filaments, indicated by the six dashed red vertical lines on the left side of Fig.~\ref{fig:imped_device_schematic}. These filaments are biased negatively with respect to the grounded chamber wall, and the discharge voltage can be varied over a broad range to control plasma conditions. Plasma initiation is achieved by applying a potential of \SI{-80}{\volt} to the heated filaments relative to the chamber. The principal operational and plasma parameters relevant to the present study are summarized in Table~\ref{tab:simParameters}. The experiments reported here are carried out at $R_m = 35$ and a neutral pressure of $5 \times 10^{-5}$~mbar. The magnetic field strength is systematically varied from 600~G to 1000~G in order to investigate its influence on plasma behavior. Within this range, the ion cyclotron frequency spans 22.8--38~kHz, while the ion-neutral collision frequency remains approximately constant at $\sim 1.18$~kHz. A more detailed description of the experimental configuration and diagnostics can be found in Refs.~\cite{bose_2015inverse, Bose_2015_RSI}.

\begin{table}[t]
\caption{Operational and plasma parameters in the IMPED device.}
\label{tab:simParameters}
\centering
\begin{tabular}{@{} l c @{}}
\toprule
\textbf{Parameter} & \textbf{Range} \\
\midrule
Magnetic field, $B_m$ (G) & $100$--$1200$ \\
Operating pressure (Ar) (mbar) & $2\times10^{-5}$--$10^{-3}$ \\
Peak electron density, $n_e$ (cm$^{-3}$) & $10^{9}$--$8\times10^{11}$ \\
Electron temperature, $T_e$ (eV) & $1$--$5$ \\
Ion temperature, $T_i$ (eV) & $0.1$ \\
Debye length, $\lambda_{De}$ (mm) & $0.02$--$0.47$ \\
Electron gyroradius, $r_{Le}$ (mm) & $0.04$--$0.47$ \\
Ion gyroradius, $r_{Li}$ (mm) & $1.7$--$20.0$ \\
Ion-neutral mfp, $\lambda_{in}$ (cm) & $3.1$--$154$ \\
Electron plasma frequency, $f_{pe}$ (GHz) & $0.3$--$8$ \\
Electron gyrofrequency, $\Omega_{ce}$ (GHz) & $0.3$--$3.4$ \\
Ion gyrofrequency, $\Omega_{ci}$ (kHz) & $3.8$--$46.1$ \\
Electron-neutral collision freq., $\nu_{en}$ (kHz) & $24.3$--$1358$ \\
Ion-neutral collision freq., $\nu_{in}$ (kHz) & $0.31$--$15.9$ \\
Electron-ion collision freq. (MHz) & $0.6$--$800$ \\
\bottomrule
\end{tabular}
\end{table}

\paragraph{\textbf{Diagnostics:}} Measurements of the equilibrium plasma parameters ($n,~T_e,~\phi_p$) are carried out using the I-V characteristics obtained from a single Langmuir probe (SLP), which is swept between \SI{-100}{\volt} and \SI{+20}{\volt} at a ramp frequency of \SI{10}{\hertz}. The probe has a diameter of \SI{0.5}{\milli\meter} and a length of \SI{4.2}{\milli\meter}, and is installed at port F (Figure~\ref{fig:probe_configaration}). To study wave properties such as the poloidal and radial wavenumbers ($k_\theta$, $k_r$), different configurations of cylindrical Langmuir probes are used, including radially and azimuthally separated multi-tip arrangements. Fluctuation measurements are performed at port C using a three-tip probe (length = \SI{5}{\milli\meter}, diameter = \SI{0.5}{\milli\meter}), which simultaneously records two density fluctuation signals ($\tilde{n}$) and one floating potential fluctuation signal ($\tilde{\phi}_f$) at the same radial location but different poloidal positions. A four-tip probe installed at port A, aligned along the $r$-$\theta$ direction, is used to verify phase relationships and mode number estimation. At port B, another four-tip probe arranged in the $r$-$\theta$ plane is used as a Reynolds stress probe to obtain the $\langle \tilde{v}_r \tilde{v}_\theta \rangle$ at radial location. The fluctuating velocity components are estimated from the floating potential fluctuation measurements using the $\mathbf{E}\times\mathbf{B}$ drift relation ($v_x=\tilde{E_x}/B_m$, x is coordinate long which $\tilde{E_x}$ is calculated), where $\tilde{v}_r = -\frac{1}{B}\frac{\partial \tilde{\phi}_f}{\partial \theta}$ and $\tilde{v}_\theta = \frac{1}{B}\frac{\partial \tilde{\phi}_f}{\partial r}$. These spatial derivatives are approximated using the potential differences between spatially separated probe tips in the poloidal and radial directions, respectively. The poloidal wavenumber ($k_\theta$) is estimated using spatially separated probe pairs, namely between $C_1$ and $C_2$ ($\Delta y = 6.4$ mm), $A_3$ and $A_4$ ($\Delta y = 10$ mm), and $B_3$ and $B_4$ ($\Delta y = 8$ mm) from port C, port A, and port B, respectively, using both density and potential fluctuation signals. Due to the axial symmetry and uniform nature of the plasma column, measurements at different axial ports can be compared at the same radial position, allowing cross-validation of both mean and fluctuation quantities. All probes are mounted on a digitally controlled servo positioning system. Radial scans are performed using a closed-loop PID-controlled DC servo drive with a positioning accuracy of \SI{0.2}{\milli\meter} \cite{Roy2023_FED, patel2025simultaneous}. However, due to alignment and calibration uncertainties between different probe assemblies, an error of up to 2 mm may be present when comparing measurements at the same nominal radial location. Data acquisition is carried out using a high-speed PXIe-based system at a sampling rate of 1 MHz. Each measurement consists of a 1-second time series, corresponding to $10^6$ data points, ensuring good statistical reliability. 

\begin{figure*}[!hbt]
\centering
\includegraphics[width=0.9\linewidth]{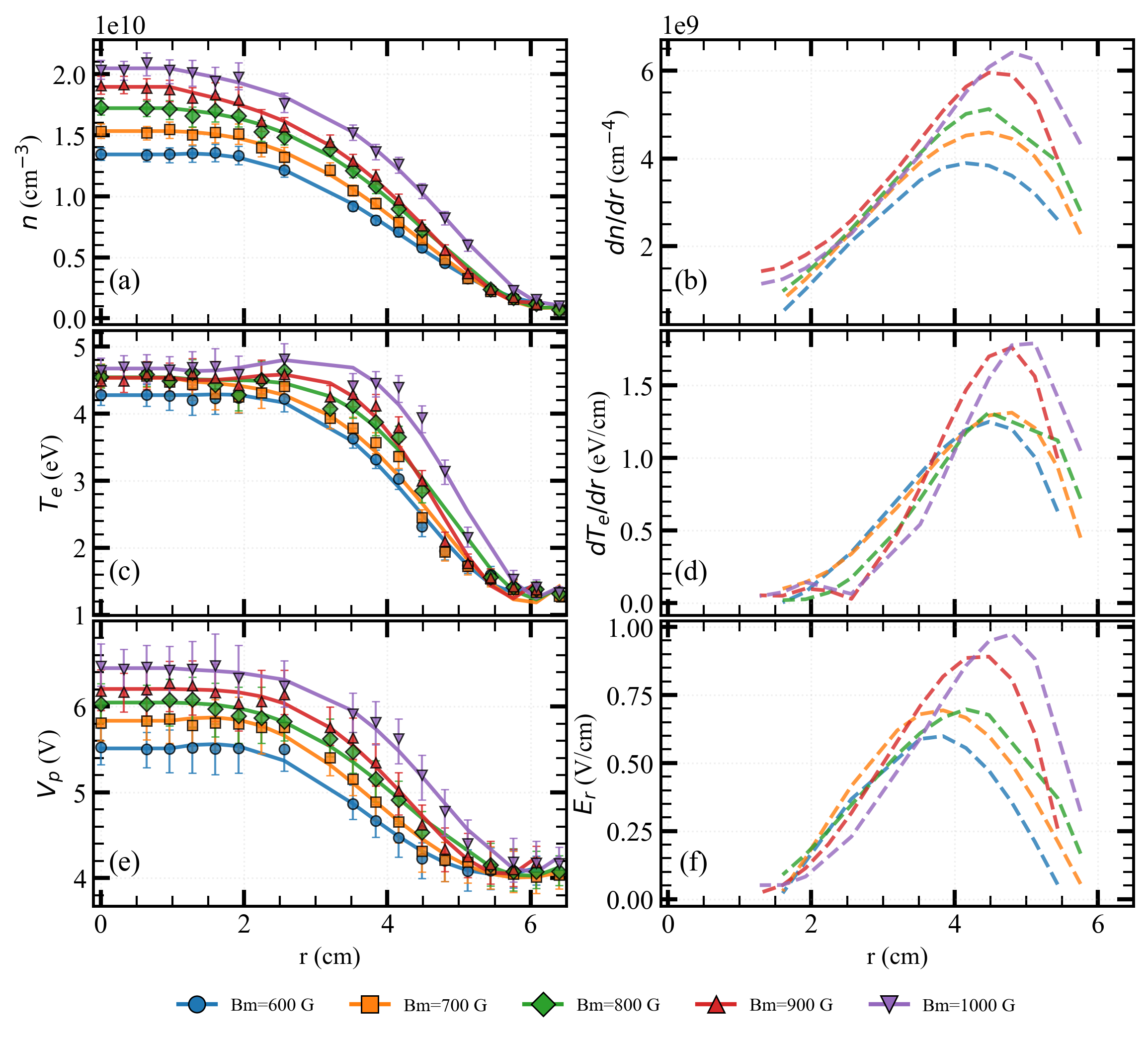} 
\caption{Parametric evolution of plasma profiles with increasing magnetic field $B_m$ from 600 G to 1000 G. (a-b) Electron density ($n$) profile and its radial gradient. (c-d) Electron temperature ($T_e$) profile and its radial gradient. (e-f) Plasma potential ($V_p$) profile and its radial gradient.}    
\label{fig:DC_mean_profile}  
\end{figure*}

\paragraph{\textbf{Data Analysis Methods:}} Plasma parameters such as $n$, $T_e$, and $\phi_p$ are obtained at each radial position by averaging the I-V characteristics over three voltage sweep cycles, followed by fitting of the averaged curve. The uncertainty in the DC profiles is estimated from the standard deviation across repeated measurements. Fluctuation data of density and potential are acquired for different values of $B_m$ (ranging from 600 G to 1000 G) at a fixed working pressure of $5\times 10^{-5}~mbar$. Earlier studies \cite{perks_2022impact, oldenburger_2012dynamics, roy2025_experimental, Roy_RT-KH_10.1063/5.0291547, karmakar2026zonal} have shown that, under similar experimental conditions, temperature fluctuations are comparatively weak in relation to density and potential fluctuations. Therefore, ion saturation current ($\tilde{I}_{sat}$) and floating potential ($\tilde{\phi}_f$) are used as representative measures of density and plasma potential fluctuations, respectively. To examine the spectral features of the fluctuations, fast Fourier transforms (FFTs) \cite{smith_2007fast} are performed using 16384 data points, resulting in a frequency resolution of \SI{61}{\hertz}. To improve statistical reliability, the spectra are averaged over 122 overlapping segments with 50\% overlap, and a Hanning window is applied to reduce spectral leakage effects. The relative contribution of fluctuations within a selected frequency range is quantified using the fractional power \cite{fujisawa2007causal, fujisawa2008experimental}, defined as $P_{\mathrm{band}} = \left( \int_{f_1}^{f_2} P(f)\,df \right) \big/ \left( \int P(f)\,df \right)$, where $P(f)$ denotes the power spectral density within full Nyquist frequency band. In addition, cross-correlation between two fluctuating signals, such as $\tilde{v}_r$ and $\tilde{v}_\theta$, is calculated to evaluate their correlation coefficient, and is expressed as $C_{xy}(\tau) = \langle x(t)\,y(t+\tau) \rangle$. To investigate nonlinear interactions among different modes, higher-order spectral techniques \cite{kim_2007_digital}, including bi-spectrum and bi-coherence analysis, are employed, which provide direct evidence of three-wave coupling processes. Furthermore, the spatio-temporal spectrum $S(k,\omega)$ is obtained using signals from poloidally separated probes, allowing simultaneous estimation of frequency and wavenumber and enabling identification of dominant modes in both spectral and spatial domains.

\begin{figure*}[]
\centering
\includegraphics[width=0.95\columnwidth]{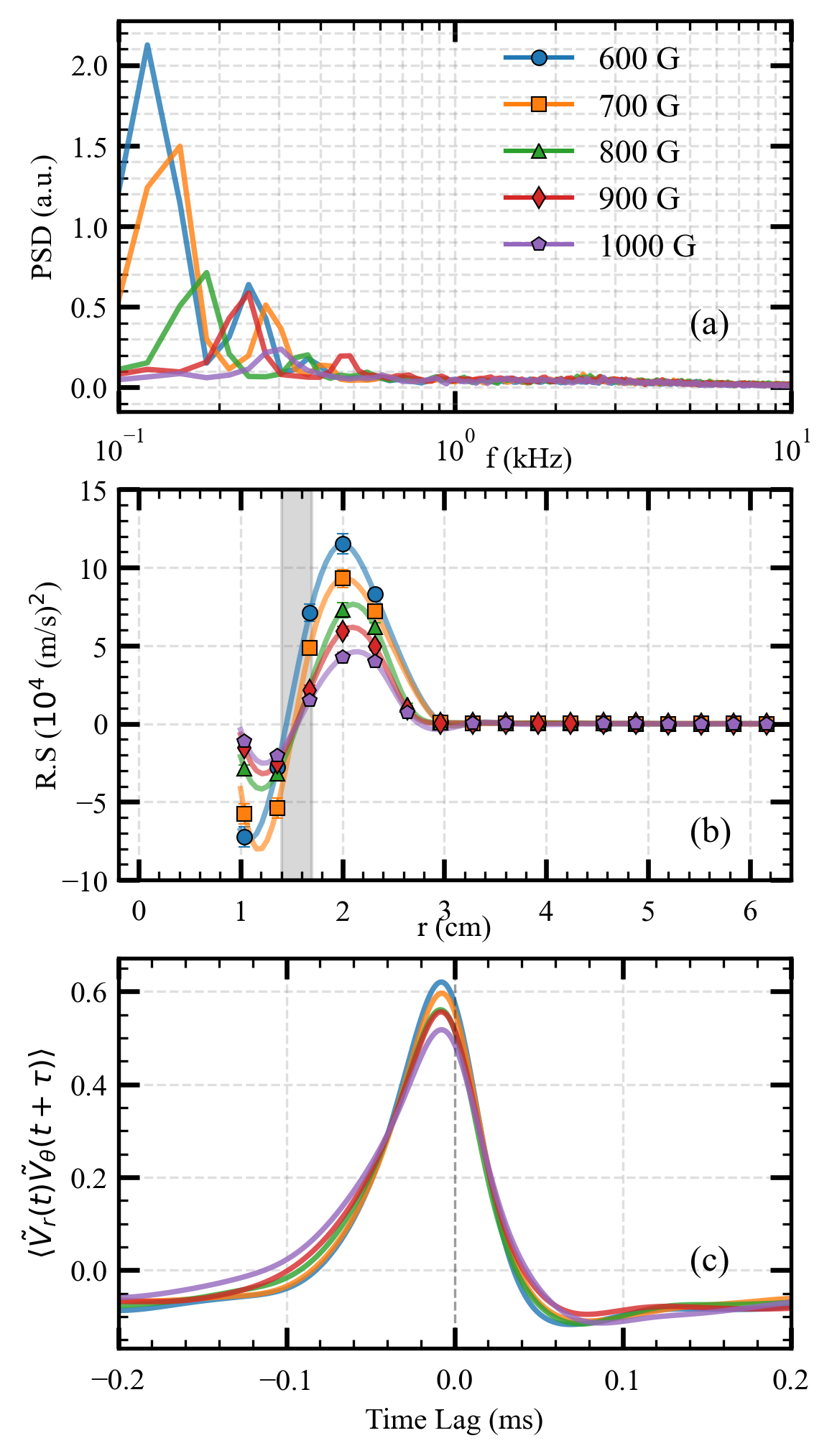}
\includegraphics[width=0.93\columnwidth]{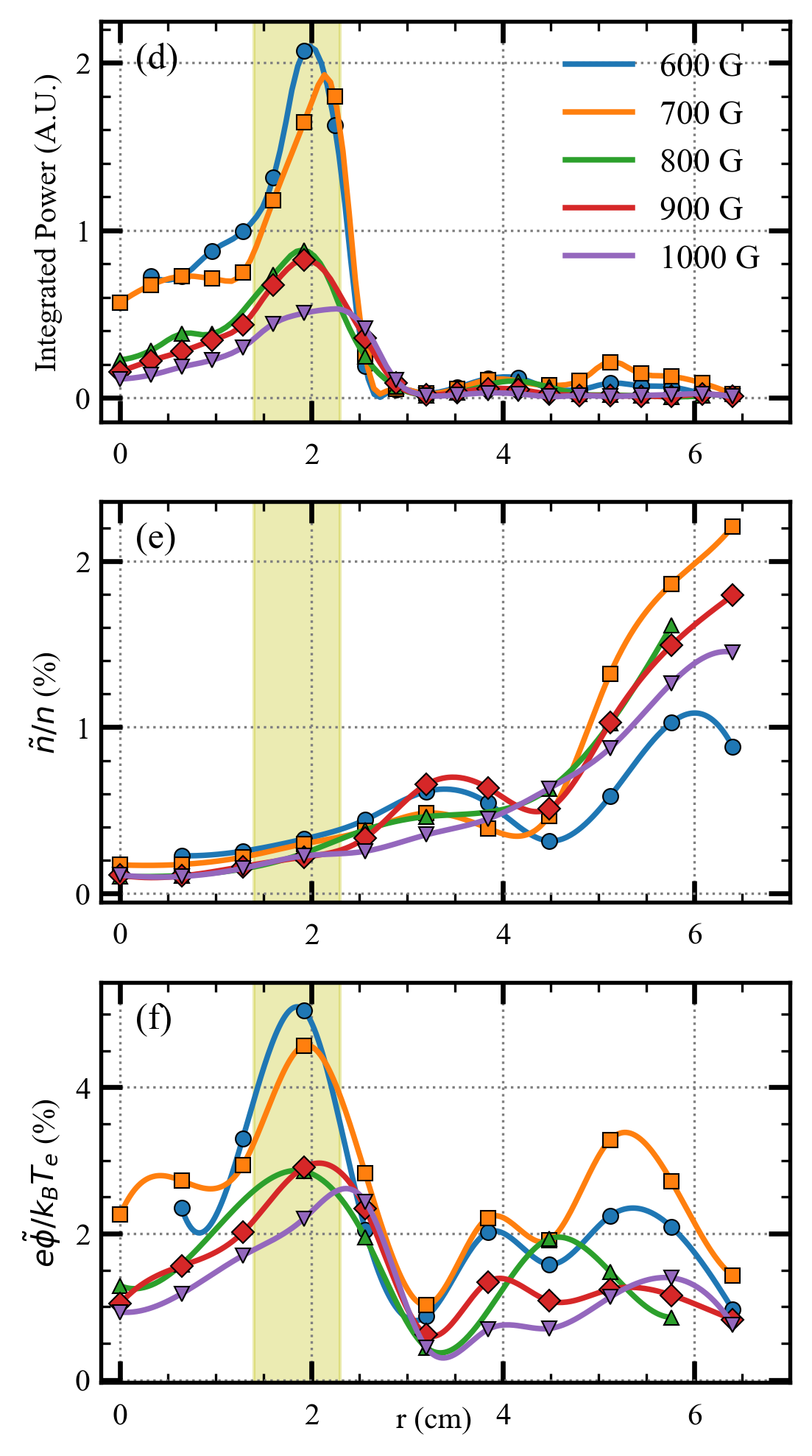} 
\caption{(a) Auto-power spectrum of the floating potential fluctuation ($\tilde{\phi}_f$) at $r = 1.6$ cm. (b) Radial profile of Reynolds stress (The region of steep gradient is highlighted in gray). (c) Cross-correlation between radial and poloidal velocity fluctuations ($\tilde{v}_r$ and $\tilde{v}_\theta$) at 1.6 cm. (d) Radial variation of integrated power in the 0.1-1 kHz frequency range. (e) Radial profile of normalized density fluctuations ($\tilde{n}/n$). (f) Radial profile of normalized potential fluctuations ($\tilde{\phi}_f/T_e$).}   
\label{fig:ZF_identification}  
\end{figure*}

\begin{figure}[]
\centering
\includegraphics[width=1\columnwidth]{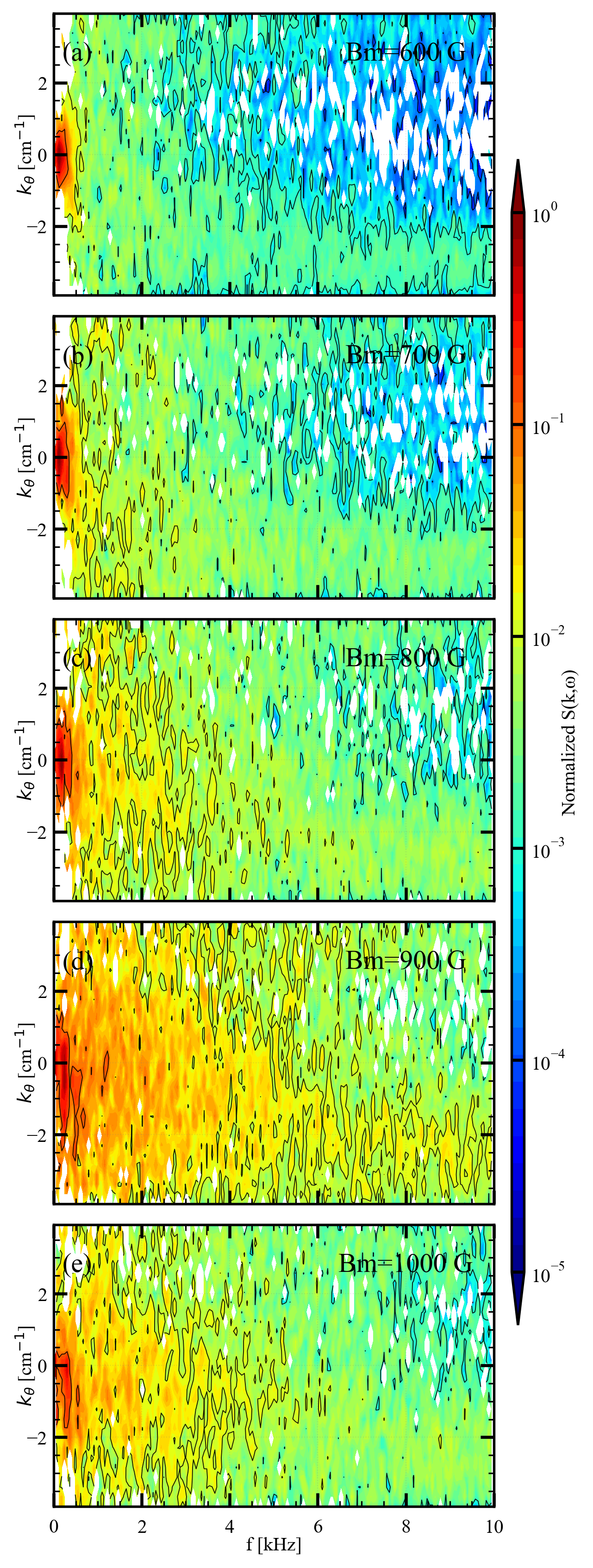} 
\caption{Poloidal wavenumber ($k_\theta$) spectra of the potential fluctuations $\tilde{\phi}_f$ at $r = 1.6$ cm for different magnetic field $B_m$.}   
\label{fig:skw_ZF}  
\end{figure}

\begin{figure*}[]
\centering
\includegraphics[width=0.95\columnwidth]{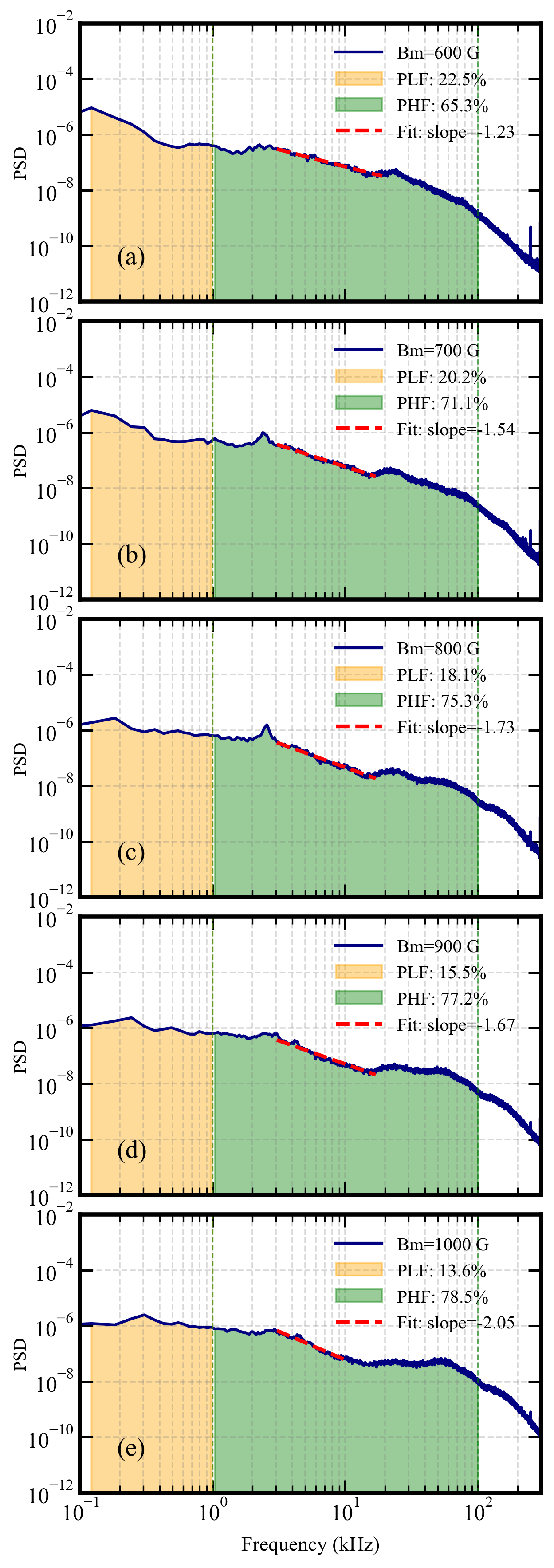} 
\includegraphics[width=0.95\columnwidth]{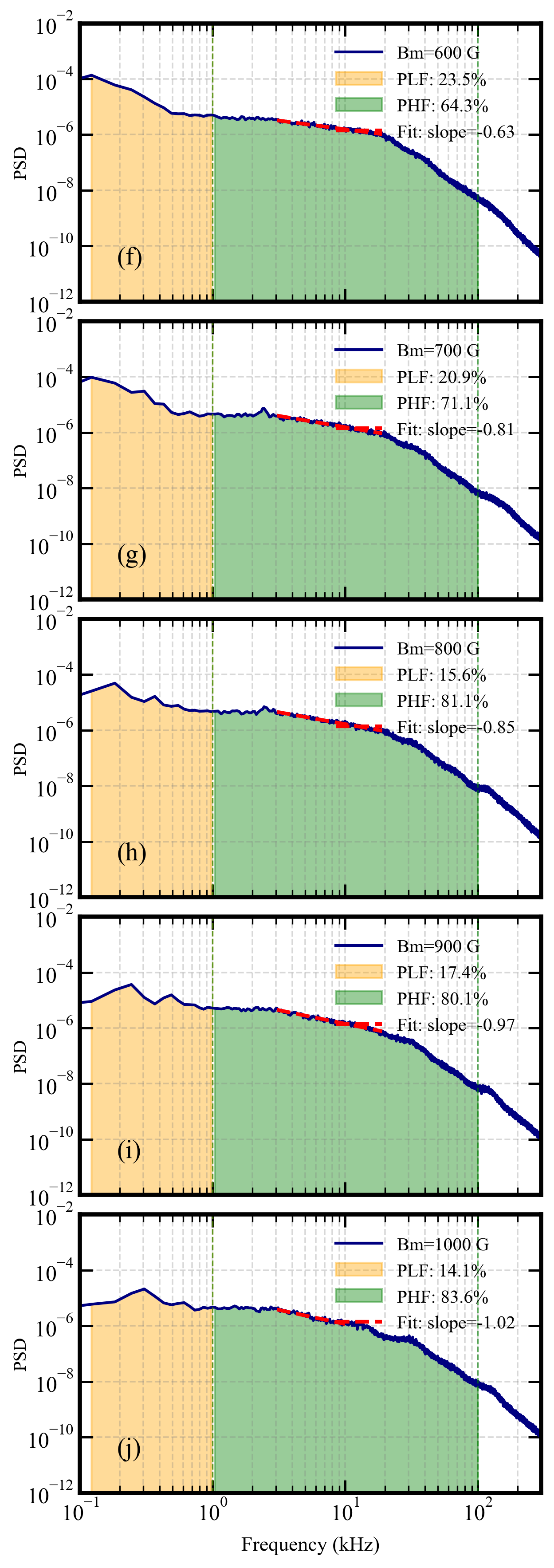}
\caption{Auto-power spectra of (a-e) density fluctuations ($\tilde{n}$) and (f-j) potential fluctuations ($\tilde{\phi}_f$) for $B_m = 600$-1000 G. Shaded regions denote low-frequency (0.1-1 kHz) and high-frequency (1-100 kHz) bands. A decrease in low-frequency power and an increase in high-frequency power with $B_m$ indicate spectral redistribution toward smaller scales.}   
\label{fig:Power_ratio_density_potential_PSD_with_Bm}  
\end{figure*}
\begin{figure}[]
\centering
\includegraphics[width=0.9\columnwidth]{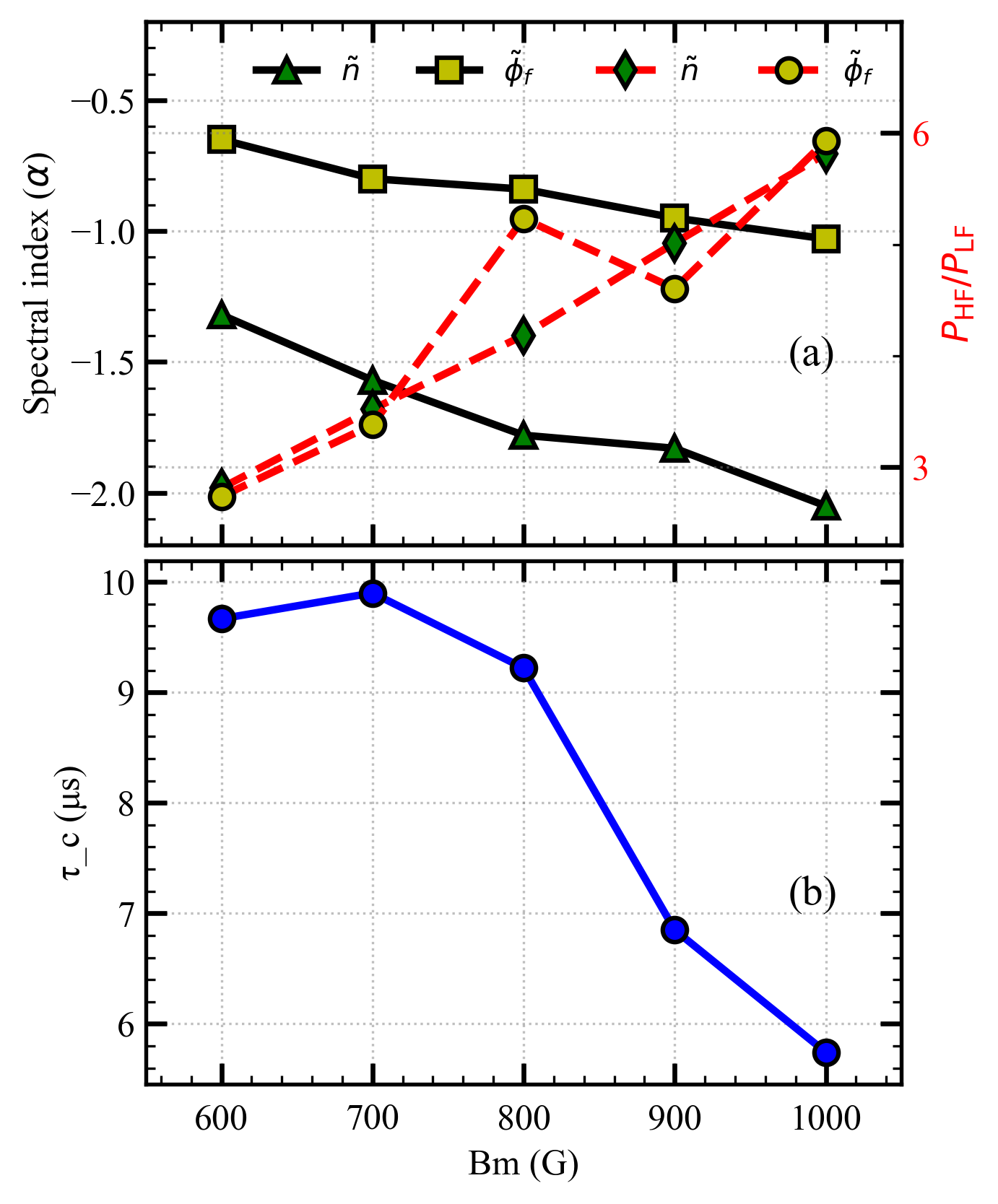} 
\caption{(a) Spectral index ($\alpha$) and power ratio ($P_{\mathrm{HF}}/P_{\mathrm{LF}}$) for density ($\tilde{n}$) and floating potential ($\tilde{\phi}_f$) as a function of magnetic field $B_m$, measured at $r=1.6$ cm (location of maximum Reynolds stress gradient). 
(b) Corresponding variation of decorrelation time ($\tau_c$) with $B_m$.}   
\label{fig:Power_ratio_with_Bm}  
\end{figure}
 
\section{Mean radial profile}
\noindent
The variation of mean plasma parameters with increasing magnetic field provides important insights into the underlying turbulence phenomena. Figure \ref{fig:DC_mean_profile} shows the radial profiles of plasma density, plasma temperature, and plasma potential, along with their corresponding gradients \cite{merlino_2007understanding,bose_2017langmuir,karmakar2026zonal, roy2025_experimental}, for a fixed magnetic field ratio $R_m=35$ and neutral pressure $5\times10^{-5}$ mbar, while the axial magnetic field $B_m$ is varied from 600 G to 1000 G. The density profile remains relatively flat in the core region ($r \lesssim 2.5$ cm) and then decreases sharply toward the edge, forming a well-defined gradient region between $r \sim 3$-5 cm (figure \ref{fig:DC_mean_profile}(a-b)). As $B_m$ increases, the overall density level increases systematically by about $40$-$50\%$ from 600 G to 1000 G, indicating improved confinement. The corresponding density gradient $dn/dr$ exhibits a peak around $r \sim 4$-4.5 cm, and its magnitude increases significantly (by nearly a factor of $\sim 1.5$-2), while the peak location shifts slightly outward with increasing $B_m$. A similar trend is observed for the electron temperature profile, which is nearly uniform in the core ($\sim 4$-5 eV) and drops toward the edge. With increasing $B_m$, the temperature increases moderately (by $\sim 10$-15\%) and its gradient $dT_e/dr$ becomes steeper, with the peak again located near $r \sim 4$-5 cm (figure \ref{fig:DC_mean_profile}(c-d)). The plasma potential profile also follows the same trend, showing a gradual decrease toward the edge, while the radial electric field $E_r$ increases in magnitude with $B_m$ (maximum value of $\sim 60$--$100$ V/m), with its peak shifting slightly outward and increasing by $\sim 30$--40\% (figure \ref{fig:DC_mean_profile}(e-f)). The corresponding $\mathbf{E}\times\mathbf{B}$ drift velocity, estimated as $V_{E\times B} = E_r/B_m$, reaches a maximum value of $\sim 1000$ m/s across the range of magnetic fields. Although $E_r$ increases with $B_m$, the increase in magnetic field partially compensates this effect, resulting in a moderate change in the mean flow velocity. However, the radial shear in the mean flow, given by $dV_{E\times B}/dr$, becomes stronger due to the steepening of the $E_r$ profile, particularly near the gradient region. This enhanced shear can influence turbulence dynamics by modifying eddy structures and contributing to the decorrelation of fluctuations \cite{Terry_2000suppression}. The steepening of gradients can be understood as a consequence of reduced cross-field transport at higher magnetic field, which limits radial diffusion and allows sharper profile formation. Moreover, the normalized density gradient scale length, defined as $L_n = [(1/n)(dn/dr)]^{-1}$ \cite{Roy_RT-KH_10.1063/5.0291547}, is also modified with increasing magnetic field. At $r \sim 4$ cm (considering finite density gradient at all $B_m$), $L_n$ increases from $\sim 2$ cm at 600 G to $\sim 2.5$ cm at 1000 G, indicating a broadening of the effective gradient scale. The ion sound Larmor radius, given by $\rho_s = \sqrt{T_e m_i}/(eB)$ (for argon ions with assumed $T_i = 0.01$ eV), decreases from $\sim 1.92$ cm to $\sim 1.32$ cm as $B_m$ increases from 600 G to 1000 G due to the stronger magnetic field. As a result, the ratio $\rho_s/L_n$ decreases from $\sim 0.96$ to $\sim 0.55$. This reduction implies that the characteristic turbulence scale becomes smaller relative to the background gradient scale length. Such a change triggers more turbulence drive and smaller-scale fluctuations \cite{burin_2005transition}. The corresponding fluctuation characteristics are discussed in the following section.

\section{Evolution of turbulence spectra}
\subsection{Low frequency fluctuations}
As discussed in the previous section, the increase in magnetic field $B_m$ leads to a reduction in the ratio $\rho_s/L_n$ at the location of peak density gradient, which is expected to influence the nature of turbulence and its dominant scales. To understand how this affects the generation of large-scale structures, we first examine the role of fluctuation-driven momentum transport through the Reynolds stress \cite{Saikat_2018simultaneous, diamond2005zonal}. In a turbulent plasma, the velocity field can be decomposed as $v = \langle v \rangle + \tilde{v}$, where the nonlinear interaction of the fluctuating components gives rise to a net momentum flux, $\langle \tilde{v}_r \tilde{v}_\theta \rangle$, known as the Reynolds stress. The gradient of Reynolds stress is responsible for driving zonal flows \cite{itoh2006physics, Saikat_2018simultaneous}. Therefore, analyzing its radial profile provides a measure of the driving source for such large-scale flows. The Reynolds stress profiles for different $B_m$ are shown in Fig.~\ref{fig:ZF_identification}(b), where a strong gradient is observed in the radial range of $1.4$--$1.8$ cm. With increasing $B_m$, this gradient decreases systematically, indicating a reduction in the drive available for zonal flow generation. The reduction in Reynolds stress with increasing $B_m$ can be understood from Fig.~\ref{fig:ZF_identification}(c). It is observed that the peak cross-correlation between $\tilde{v}_r$ and $\tilde{v}_\theta$ decreases systematically as $B_m$ increases. This indicates a loss of correlation between the radial and poloidal fluctuating velocity components. Since Reynolds stress is directly related to the correlated product $\langle \tilde{v}_r \tilde{v}_\theta \rangle$, a reduction in this correlation leads to a weakening of the Reynolds stress drive. To investigate how this reduction in drive affects the fluctuations, we analyze the auto-power spectrum of the floating potential fluctuation $\tilde{\phi}_f$ at $r = 1.6$ cm, which lies within the region of maximum Reynolds stress gradient (Fig.~\ref{fig:ZF_identification}(a)). The choice of this location is motivated by the expectation that zonal flows, if present, would be most driven \cite{diamond_2004review, karmakar2026control, nagashima_2008coexistence} and therefore most clearly observed near this region. The spectrum shows that the power in the low-frequency range ($0.1$--$1$ kHz) decreases with increasing $B_m$, while the peak frequency shifts from $\sim 117$ Hz to $\sim 490$ Hz. Since zonal flows are typically associated with low-frequency, large-scale potential fluctuations, this reduction in low-frequency power suggests a weakening of zonal flow oscillation \cite{nagashima_2008coexistence, fujisawa_2004identification}. To further confirm the nature of these low-frequency fluctuations, we examine their spatial characteristics in wavenumber space. For this purpose, the poloidal wavenumber spectrum ($k_\theta$) is calculated using two poloidally separated $\tilde{\phi}_f$ signals (probe tips $B_3 ~\&~ B_4$ of port B) at the same radial location (Fig.~\ref{fig:skw_ZF}). The results show that the modes corresponding to the low-frequency range indeed have $k_\theta \approx 0$, i.e., they are poloidally symmetric structures \cite{karmakar2026zonal, diamond2005zonal}, which is further confirmed using measurements from port A. At the same time, the corresponding radial wavenumber, $k_r$, remains finite ($\approx 0.6$-$1~\mathrm{cm}^{-1}$), consistent with the characteristic structure of zonal flows, for which $k_\theta = 0$ and $k_r \neq 0$. In addition to spectral identification, it is important to determine the spatial localization of these fluctuations. For this purpose, the radial variation of integrated power in the low-frequency band (0.1-1 kHz) is analyzed (Fig.~\ref{fig:ZF_identification}(d)). This identifies the region where the fluctuations are localized. The results show a clear peak in the radial range of $1.5$--$2$ cm for all values of $B_m$ with a decrease in power with increasing $B_m$, and coinciding with the region of strong Reynolds stress gradient. This spatial overlap provides further evidence that these fluctuations are driven by the local Reynolds stress gradient. Finally, to understand the nature of the fluctuations, whether it is density or potential fluctuation driven, we compare the normalized potential and density fluctuation levels (Fig.~\ref{fig:ZF_identification}(e-f)). It is observed that the normalized potential fluctuation ($\tilde{\phi}_f/T_e$) is nearly an order of magnitude larger than the normalized density fluctuation ($\tilde{n}/n$), indicating that the fluctuations are predominantly driven by potential fluctuations, as expected for zonal flows. Taken together, all these observations-low-frequency dominance, $k_\theta \approx 0$ structure, finite $k_r$, spatial localization at the Reynolds stress gradient, and dominance of potential fluctuations-consistently identify the observed low-frequency fluctuations as zonal flow-like structures. Furthermore, the systematic reduction in their amplitude with increasing $B_m$ can be directly linked to the weakening of the Reynolds stress gradient, which reduces the drive for zonal flow generation and leads to their gradual suppression \cite{saikat_2013suppression}. However, the observed reduction in low-frequency power suggests a possible redistribution of spectral energy across smaller turbulent scales, consistent with the constraint of overall energy conservation within the accessible dynamical scales of the system. This aspect is examined in the following section.

\begin{figure*}[t]
\centering
\includegraphics[width=0.95\textwidth]{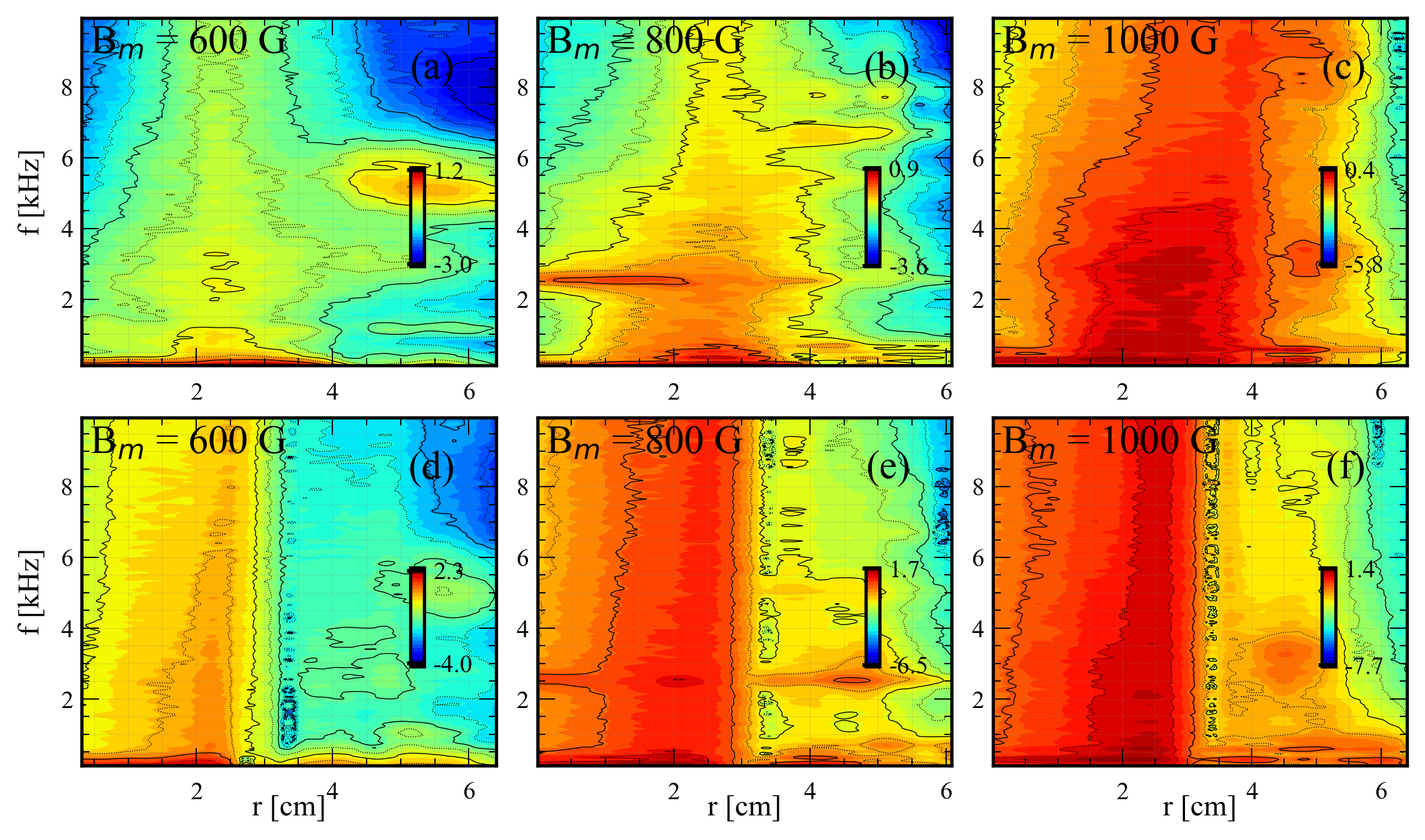} 
\caption{Radial fluctuation spectra for different magnetic fields. Panels (a)--(c) show the density fluctuation spectra, while panels (d)--(f) show the floating potential fluctuation spectra for $B_m = 600$, $800$, and $1000$~G, respectively. With increasing $B_m$, the spectra evolve towards small-scale dominated turbulent regime.}   
\label{fig:Contour_plot}  
\end{figure*}


\subsection{High frequency fluctuation}
\begin{figure}[]
\centering
\includegraphics[width=\columnwidth]{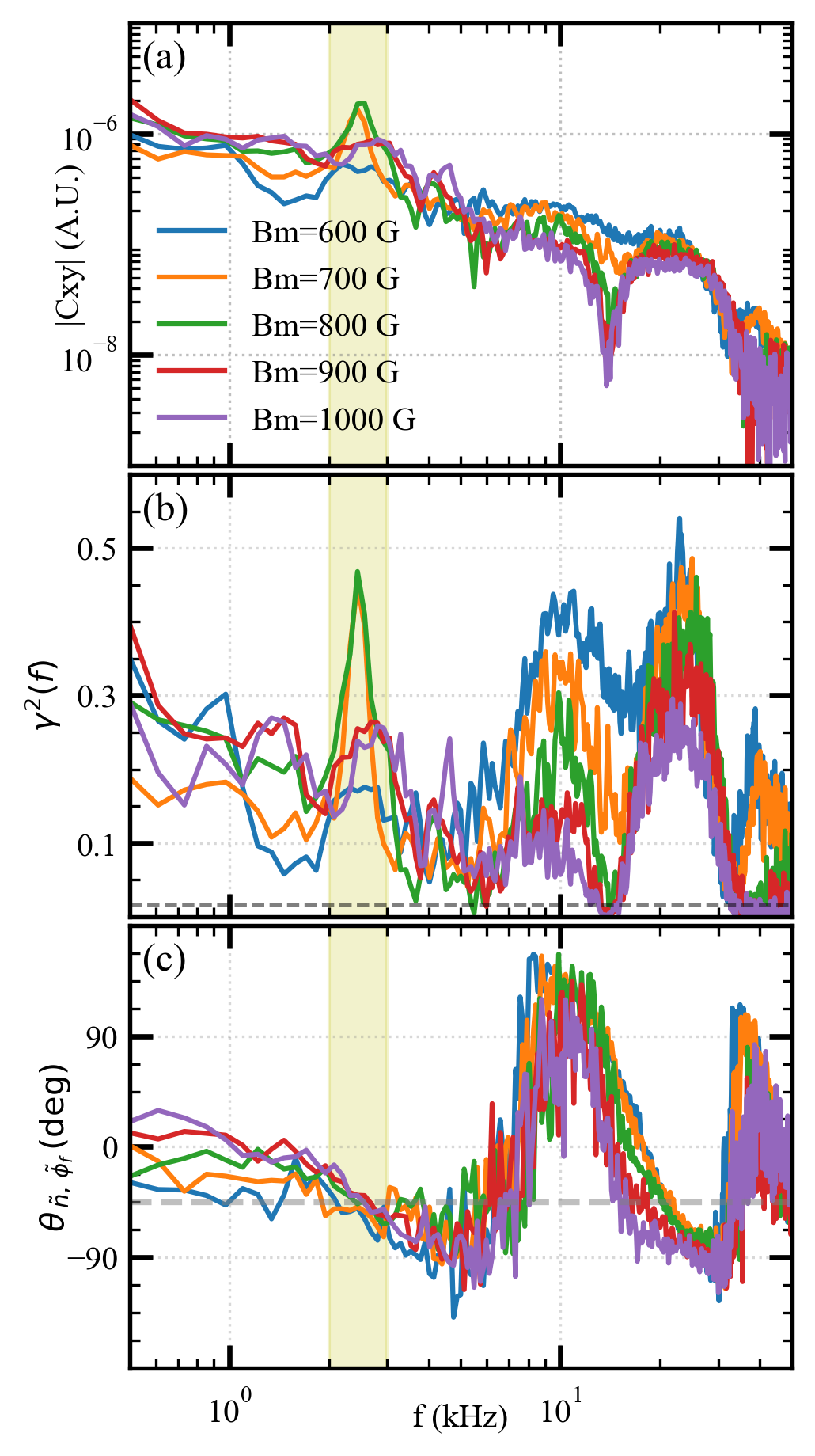} 
\caption{Variation of (a) cross-power spectrum, (b) squared coherence, and (c) cross-phase between density fluctuations $\tilde{n}$ and floating potential fluctuations $\tilde{\phi}_f$ with magnetic field $B_m$ at 1.6 cm.}   
\label{fig:CSD_coherence_phase}  
\end{figure}

\noindent
To understand the underlying nature of the observed high-frequency fluctuations and turbulence, radial fluctuation spectra of the $\tilde{n}$ and $\tilde{\phi}_f$ signals are analyzed, as shown in Fig.~\ref{fig:Contour_plot}.  Figures~\ref{fig:Contour_plot}(a--c) correspond to the density fluctuation spectra, while Figs.~\ref{fig:Contour_plot}(d--f) represent the floating potential fluctuation spectra for increasing magnetic field strength $B_m$. Although only three representative cases are shown here, similar trends are observed consistently over the entire range of $B_m$ from 600 G to 1000 G. At lower $B_m$, the fluctuation power is predominantly confined below 1 kHz, where signatures of zonal flows (ZFs) are observed, indicating the presence of relatively coherent large-scale structures. As $B_m$ increases, the spectral power progressively broadens toward higher frequencies, corresponding to comparatively smaller-scale fluctuations. This spectral broadening is followed by a transition from a coherent fluctuation regime to a more broadband turbulent state, indicating redistribution of turbulent energy from large-scale coherent modes to smaller-scale turbulent fluctuations. Figure~\ref{fig:CSD_coherence_phase}(a) shows the cross-power spectrum of different $B_m$. A weak coherent peak is observed around $f \sim 2.3$ kHz superimposed on a broadband turbulent background in the intermediate magnetic field range of 700-800 G. This frequency lies close to the expected theoretical drift frequency range. Using the measured plasma parameters ($T_e \approx 2.7$ eV, $L_n \approx 1.8$ cm, and $k_\theta \approx -0.2$ cm$^{-1}$, corresponding to the electron diamagnetic drift direction) at 4.5 cm, the Doppler-corrected drift frequency with respect to the background $\mathbf{E}\times\mathbf{B}$ velocity \cite{roy2025_experimental, karmakar2026zonal}, is estimated to be $f_{*} \sim 3.6$ kHz, which is close to the observed peak. This indicates that the coherent component of the observed turbulence is a drift mode. The squared coherence spectrum has been shown in Fig.~\ref{fig:CSD_coherence_phase}(b) where a finite level of coherence is observed in the same frequency range, that is above the spectral noise floor ($\sim$ 0.01), indicating that density and potential fluctuations are correlated at these scales. The corresponding cross-phase spectrum in Fig.\ref{fig:CSD_coherence_phase}~(c) shows that the phase difference between $\tilde{n}$ and $\tilde{\phi}_f$ lies in the range of $40^\circ$-$50^\circ$ around the peak frequency. Such a finite phase lag is one of the characteristic features of resistive drift waves, where collisions introduce a phase shift between density and potential fluctuations. However, this coherent feature is not clearly distinguishable at higher ($B_m = 900-1000~ G$) magnetic fields. This is mostly because, at this higher magnetic field, the spectrum goes into fully turbulent state. As the magnetic field is increased from 600 G to 1000 G, a gradual change in the spectral characteristics is observed. The distinct coherent peak seen at lower magnetic field becomes less pronounced and eventually transits into a fully developed broadband turbulent spectrum, consistent with the observation seen in the evolution of fluctuation spectra (Figure\ref{fig:Contour_plot}). The cross-power decreases in the low-frequency range, while the coherence level also reduces, indicating a loss of correlation between density and potential fluctuations. This transition is consistent with the decreasing $\rho_s/L_n$ ratio with increasing $B_m$, which shifts the spectral power towards the smaller scales. Thus, these observations suggest that at a lower magnetic field, the fluctuations retain signatures of drift-wave dynamics with a weak coherent component, finite cross-phase, and frequency close to the drift frequency. However, with increasing $B_m$, the system evolves toward a more fully developed turbulent state. Nevertheless, the persistence of drift-wave-like characteristics in the phase relation, propagation direction, and frequency scaling indicates that the observed turbulence is most likely drift-wave-driven, with its nature gradually modified by increasing magnetic field.

\subsection{Spectral Reorganization of Turbulence}
\noindent
To understand how turbulence evolves with increasing magnetic field, we examine the spectral characteristics of both density and potential fluctuations over a wide frequency range. Figures~\ref{fig:Power_ratio_density_potential_PSD_with_Bm}(a-e) and \ref{fig:Power_ratio_density_potential_PSD_with_Bm}(f-j) show the auto-power spectra of normalized density and potential fluctuations, respectively, for $B_m$ varying from 600 G to 1000 G. At lower magnetic field, the spectra exhibit relatively higher power in the low-frequency (LF) range ($0.1$-$1$ kHz), which is generally associated with large-scale ZFs like structures. As $B_m$ is increased, a gradual reduction in power within this LF band is observed, while the spectral power at higher frequencies ($1$-$100$ kHz) increases. To quantify this change, we calculate the fractional power in those two frequency bands \cite{fujisawa_2008review, fujisawa2007causal, fujisawa2008experimental}, defined as $P_{\mathrm{band}} = \left( \int_{f_1}^{f_2} P(f)\,df \right) \big/ \left( \int P(f)\,df \right)$, where $P(f)$ is the power spectral density upto nyquist frequency limit and ($f_1$-$f_2$) represent the frequency limits of the chosen band. Using this definition, it is found that the fractional power in the LF band decreases systematically with increasing $B_m$, while the fractional power in the high-frequency (HF) band increases. The ratio of HF to LF power also increases monotonically; the ratio becomes maximum at 1000 G (figure \ref{fig:Power_ratio_with_Bm}a), indicating a progressive redistribution of fluctuation power across scales. This is also evident from Fig.~\ref{fig:Contour_plot}, where the system is observed to evolve from a coherent state to a turbulent state. This observed trend suggests that, with increasing magnetic field, the contribution of large-scale, low-frequency fluctuations becomes weaker, while smaller-scale, higher-frequency fluctuations become more dominant.  From the $k_\theta$ spectra (figure \ref{fig:skw_ZF}), it can also be seen that, with increasing $B_m$, finite spectral power is being reorganized among higher scales (large $k_\theta$) (at 600 G, it is dominated by low $k_\theta$ whereas, at 900-1000 G, it is dominated by large $k_\theta$) Such a change in turbulent scale along with localization of evolution turbulence (The dominant spectral power is distributed over a broad frequency range at high $B_m$ within the radial region of $r = 1.5$--$3~\mathrm{cm}$) can be understood in connection with the reduction of Reynolds stress drive discussed earlier, which weakens the mechanism responsible for sustaining large-scale coherent structures. Consistently, the cross-correlation between $\tilde{v}_r$ and  $\tilde{v}_\theta$ also decreases (figure \ref{fig:ZF_identification}c) with increasing $B_m$, indicating a loss of phase coherence between the fluctuating velocity components, which actually causes the reduction of Reynolds stress drive and a reduced efficiency of momentum transfer to large scale structures like ZFs. At the same time, the increase in mean flow shear with $B_m$ (figure \ref{fig:DC_mean_profile}f) can lead to tilting of turbulent eddies, thereby favoring the formation of smaller-scale structures. The analysis of the spectral slope of the auto power spectra (figure \ref{fig:Power_ratio_density_potential_PSD_with_Bm}) provides further insight into the evolution of turbulence with increasing magnetic field. The spectral index is estimated from a linear fit to the log--log power spectrum over the frequency range of $3$--$18~\mathrm{kHz}$, where an approximately constant slope, $\mathrm{d}(\log P)/\mathrm{d}(\log f)$, is observed, excluding the low-frequency coherent features and high-frequency noise-dominated regions (see Fig.~\ref{fig:Power_ratio_density_potential_PSD_with_Bm}(a)). It is found that the spectral index systematically increases with $B_m$ for both density (from $-1.23$ to $-2.05$) and potential fluctuations (from $-0.8$ to $-1.02$), indicating a progressive steepening of the spectra (figure \ref{fig:Power_ratio_with_Bm}a). This trend reflects a relatively faster decay of spectral power toward higher frequencies and suggests a redistribution of spectral energy toward smaller scales. Consistent with this observation,  Figure~\ref{fig:Power_ratio_with_Bm}(b) shows a monotonic decrease in the turbulent de-correlation time $\tau_c$ within the 1-300 kHz range as $B_m$ increases, indicating faster turbulent dynamics and shorter-lived structures. Importantly, similar variations in both the fractional power distribution and the spectral slope are observed not only at a single radial location but also across the region of maximum profile gradients, suggesting that the change in turbulence characteristics is not localized but extends throughout the plasma column. Taken together, these results point to a systematic reorganization of turbulence with increasing magnetic field, where the relative contribution gradually shifts from large-scale, low-frequency structures to smaller-scale, higher-frequency fluctuations. This transition appears to be driven by the combined effects of reduced Reynolds stress, loss of fluctuation coherence, and increased mean flow shear, which together modify the way energy is distributed across turbulent scales.

\section{Discussion and Conclusion}
\noindent
The results of the present study bring out a clear picture of how turbulence evolves as a function of magnetic field and how the underlying regulation mechanism is modified in this process. The analysis of mean profiles shows that higher $B_m$ leads to an increase in density and temperature along with modified gradient scale lengths, and a reduction in the ratio $\rho_s/L_n$, which sets the relative scale of turbulence. In this background, the fluctuation analysis reveals that the Reynolds stress gradient, which acts as the primary drive for zonal flow generation \cite{diamond_2004review, fujisawa_2008review, BDT_Theory_1990influence}, decreases systematically with increasing $B_m$. This reduction in drive is further supported by the observed decrease in cross-correlation between $\tilde{v}_r$ and $\tilde{v}_\theta$, indicating a loss of phase coherence among fluctuating velocity components. As a result, the ability of the system to sustain large-scale coherent structures such as zonal flows is weakened. This change is clearly reflected in the spectral behavior of fluctuations, where a systematic reduction in low-frequency power and a corresponding increase in high-frequency power are observed with increasing $B_m$. The radial power spectra (figure \ref{fig:Contour_plot}), fractional power analysis and spectral slope variation (figure \ref{fig:Power_ratio_density_potential_PSD_with_Bm}) together show that the contribution from large-scale, low-frequency fluctuations decreases, while smaller-scale, higher-frequency fluctuations become more dominant. At the same time, the increase in mean flow shear also contributes to this process by distorting and breaking larger eddies into smaller structures \cite{Terry_2000suppression,BDT_Theory_1990influence}. Importantly, these trends are observed consistently across the plasma column, indicating that the change in turbulence characteristics is global rather than localized. In conclusion, these results show that increasing the magnetic field leads to a reorganization of turbulence from large-scale, coherent structures to smaller-scale, less correlated fluctuations through the combined effects of reduced Reynolds stress drive, loss of fluctuation coherence, and enhanced mean flow shear. This study highlights how magnetic field can control the balance between different turbulence scales and provides insight into regimes where the usual self-regulation of turbulence is modified.

\begin{acknowledgments}
Authors wish to acknowledge Kalpesh Doshi, Jignesh Patel, Kirti Mahajan, Manisha Bhandankar, Minsha Shah, Praveenlal and Sandip Das from IPR, India for their experimental support. Abhijit Sen is grateful to the Indian National Science Academy (INSA) for the INSA Honorary Scientist position.

\end{acknowledgments}

\bibliographystyle{ieeetr}
\bibliography{References}

\end{document}